\documentclass[12pt]{article}
\begin{document}

\begin{titlepage}
\begin{center}
{\Large \bf Epistemology in Cyclic Time }

\vspace{5mm}

\end{center}

\vspace{5 mm}

\begin{center}
\textbf{Moninder Singh Modgil}

\vspace{5mm} \textit{Department of Physics}\\ \textit{Indian
Institute of Technology, Kanpur, India}

\vspace{5mm} email:msingh@iitk.ac.in
\end{center}

\vspace{3mm}

\vspace{1cm}

\noindent {\bf Abstract}

\noindent Consider the scenario, in which human civilization
undergoes periodic eras of progression and regression, and
consequently, changes in cosmological knowledge are cyclic. There
exist solutions of general theory of relativity, such as the
G\"{o}del universe, in which the cosmos is rotating. If the real
universe is indeed rotating, than this would be a reversion to
rotating universe models, used in ancient cosmological models. We
argue that such reversions in physical models would be inevitable
in a space-time in which time is having $S^1$ (circular) topology.

\vspace{1cm}

 \noindent  \textbf{KEY  WORDS:} G\"{o}del universe, Ancient Cosmologies, Epistemology

\end{titlepage}

\section{Introduction}
Evolution of events in identical cycles, has been a hall mark of
ancient philosophical systems - right from American Indians, and
Greeks, on one hand, to Asian Indians, and Chinese on the other
\cite{Jaki 1974}. In more recent times, Poincare and Nietzsche
attempted a mathematical formulation of the idea, which is now
enshrined in mathematical physics as Poincare recurrence concept
\cite{Brush 1966}. In the last quarter of twentieth century,
isolated attempts to resolve certain problems of theoretical
physics, using a cyclic concept of time have been made
\cite{Villaroel 1986, Schechter 1977} - notable among them is
Segal's cosmological model \cite{Segal 1984, Guillemin 1989},
which successfully re-interprets all observational evidence,
hitherto attributed to big-bang cosmology. In this model, universe
is closed - having topology of 3-sphere, and time cyclic. In this
space-time, light rays return to their origin, after circling the
closed universe in a single time cycle. Popper \cite{Popper 1957}
cedes, that changes in knowledge occur gradually, but being an
evolutionist, he feels that the changes began billions of years
ago, many of which are in built in the structure of our sense
organs and brains - and that this series of changes in knowledge
will continue indefinitely. However, provided one combines,
Popper's concept of gradual change in human knowledge, with
Poincare's concept of result of a series of changes being a cycle,
one is lead to epistemology in cyclic time.

If one considers the possibility that human civilization has
always been extant in this cyclic time, with number of generations
of humans being a very large integer, then one is lead to the
inescapable conclusion, that human knowledge, science and
technology, also change in identical cycles. One is lead to ask
the question, as to what is the driving force in this cyclic
changes in human knowledge. As human knowledge arises by
observation and interpretation, a function of mind and intellect,
therefore, the driving force behind cyclic change in human
knowledge, must be cyclic changes in perspicuity and purity of
human mind and intellect.

As a striking example of evolution of knowledge in cycles (within
a physical cyclic time), consider the conceptual status awarded to
nature of earth, and earth's relationship with cosmos. Prior to
Aristotle, as indicated in ancient Indian texts, earth was
considered flat, and the universe rotating daily above this flat
earth. Around about Aristotle's time the world view emerged, that
various differences in Sun's latitude observed at different
northern points could be attributed to a spherical earth. However,
the universe continued to rotate daily around the spherical earth,
in Aristotelian cosmology \cite{Aristotle 1952}. During the
generations of Kepler, Copernicus, and Galileo, the accumulated
data on movements of astronomical bodies, lead them to conclude,
that earth was not the center of universe, but was a planet, and
rotated once around the sun once every year, and this was the
cause of seasons.

Both changes, the first one around Aristotle's time, and second
one around Galileo's time, were based on the generally accepted
validity of Euclidean geometry - fundamental assumption being that
light travels in straight lines. However, geometry itself,
underwent fundamental changes during 19th century, when it was
realized, that alternatives to Euclidean geometry were possible,
by modifying its axioms \cite{Maor 1987}. Notable among them was
the work of Riemann, which laid foundation for Einstein's general
theory of relativity \cite{Einstein 1956}. One immediate result of
general relativity was that light stopped travelling in straight
lines, in minds of men, after confirmation of Eddignton's solar
eclipse experiment \cite{Wheeler 1973} on bending of star light
near sun. Since then a number of solutions to Einstein's field
equation for gravitation have been found. Einstein's original
cosmological model had a 3-sphere universe, and an infinite time,
and was called "Einstein's cylinder". Hopes of general
relativists, that Einstein's field equations determine the
universe's large scale structure uniquely were dashed to the
ground by G\"{o}del's discovery of a rotating universe model
\cite{Godel 1949}, which though had a same local distribution of
matter as Einstein's cylinder, nevertheless, differed remarkably
from it in over all properties - most notable among them being
that, G\"{o}del universe had closed time like curves, and it was
possible for an observer to access its past in principle.
G\"{o}del got around this problem of causality violation, by
denying freewill to observers. Since then, a number of
generalizations of G\"{o}del universe have been created, including
those, in which universe is charged, and a constant
electromagnetic field is present everywhere within universe
\cite{Ravel 1967}.

Many modern physicists are of the view, that mankind is near
climax of conceptual knowledge \cite{Davis 1990}. Such being the
case, a believer of cyclic time, in which human observers exist
eternally, is lead to ask the question, as to how, given the
present state of human knowledge, will the humans revert back to
the belief of universe rotating daily, above a flat stationary
earth. This is an example of issues facing Epistemology in cyclic
time. One possibility is that, through a series of artificial or
natural calamities, mankind is reduced to a state deprived of its
intellectual inheritance, and therefore, out of ignorance, such
primitive world view prevails. There exists, however, an
altogether, different, striking, possibility, which I find very
exciting, and would like to share with you.

\section{Rotating Universe models in General Relativity}
As mentioned earlier, rotating G\"{o}del universe is a solution to
Einstein's field equations. Behavior of light rays, in G\"{o}del
universe, has the remarkable property, that light rays curve back,
after travelling a finite distance perpendicular to the direction
of rotation \cite{Laurent 1981}. Thus imagine, a flat earth in
G\"{o}del cosmology, with rotation axis perpendicular to the flat
earth. A picture taken from a space-craft, above this flat earth
in G\"{o}del cosmology, will show light arriving at the camera,
only from a circular patch - light from remaining parts of flat
earth curving away from camera and never reaching it. This will
generate illusion of sphericity, for one who assumes that space-
time geometry near earth is basically Euclidean. Similarly, light
from stars such as sun, will be accessible, only from a circular
patch, on this flat earth, remaining parts of earth, being
deprived, as the light curves away, as it approaches earth. At
circumference, of this circular patch, light rays will be
tangential to flat earth, thus explaining the apparent sunrise,
sunset etc..

Further if such a rotating universe is charged, it will generate a
magnetic field which reverses, periodically, as a result of
quantum cosmological tunnelling between two counter-rotating
solutions of quantum gravity Wheeler-DeWitt equation \cite{Wheeler
1967}, which describes such a universe. I have done some
calculations \cite{Modgil 1993} which show magnitude of this
geomagnetic polarity reversal to be 157 degree, which is close to
actual value measured from paleo-magnetic rock samples. My
prediction of paleomagnetic intensity during transition periods
also matches that observed in samples. Presentations of these
detailed calculations would however, be beyond scope of this
conference and paper.  I invite reader to contact me for these
details. However, time is not cyclic in G\"{o}del universe, and
universe also is not closed.  These two objections can be overcome
by what is called as "conformal compactification", \cite{Penrose
1984} in jargon of general relativists, which make the time cyclic
and universe closed, with other properties of G\"{o}del universe,
such as rotation, and presence of closed time like curves,
remaining intact.

Needless, to say a number of unresolved issues remain, before
widespread adoption of a "quantum gravity rotating G\"{o}del
universe", takes place - which will have to be tackled one at a
time, as was the situation faced by Galileo \cite{Galileo 1967}.
Interestingly, it can be shown that notion of causality in
G\"{o}del universe is actually consistent with implications of
Einstein-Podolsky-Rosen (EPR) paradox \cite{Einstein 1935}, which
is claimed to have been experimentally verified in recent times
\cite{Aspect 1981}. I am of the opinion, that if time is cyclic
with eternal existence of civilized humans, then provided a
catastrophe, which breaks the continuity of evolution of human
knowledge, does not occur, then the quantum gravity G\"{o}del
universe must be the best description of our universe. So I find
epistemology in cyclic time indeed exciting, and welcome the
reader to share in the churning on true nature of time (cyclic or
linear) and its implications.

\end{document}